\documentclass[aps,preprint]{revtex4}%
\usepackage{amsfonts}
\usepackage{amsmath}
\usepackage{amssymb}
\usepackage{graphicx}%
\providecommand{\U}[1]{\protect\rule{.1in}{.1in}}
\newtheorem{theorem}{Theorem}

\newtheorem{corollary}[theorem]{Corollary}

\newenvironment{proof}[1][Proof]{\noindent\textbf{#1.} }{\ \rule{0.5em}{0.5em}}
\begin{document}
\preprint{HEP/123-qed}
\title[Short title for running header]{Dynamical symmetries, coherent states and nonlinear realizations: the SO(2,4) case}
\author{Andrej B. Arbuzov}
\affiliation{Bogoliubov Laboratory of Theoretical Physics, Joint Institute for Nuclear
Research, 141980 Dubna, Russian Federation}
\author{Diego Julio Cirilo-Lombardo}
\affiliation{Consejo Nacional de Investigaciones Cientificas y Tecnicas (CONICET),
Universidad de Buenos Aires, National Institute of Plasma Physics (INFIP),
Facultad de Ciencias Exactas y Naturales, Ciudad Universitaria, Buenos Aires
1428, Argentina}
\affiliation{Bogoliubov Laboratory of Theoretical Physics, Joint Institute for Nuclear
Research, 141980 Dubna, Russian Federation}
\pacs{PACS number}

\begin{abstract}
Nonlinear realizations of the $SO(4,2)$ group are discussed from the point of
view of symmetries. Dynamical symmetry breaking is introduced. One linear and
one quadratic model in curvature are constructed. Coherent states of the
Klauder-Perelomov type are defined for both cases taking into account the
coset geometry. A new spontaneous compactification mechanism is defined in the
subspace invariant under the stability subgroup. The physical implications of
the symmetry rupture in the context of non-linear realizations and direct
gauging are analyzed and briefly discussed.

\end{abstract}
\volumeyear{year}
\volumenumber{number}
\issuenumber{number}
\eid{identifier}
\date[Date text]{date}
\received[Received text]{date}

\revised[Revised text]{date}

\accepted[Accepted text]{date}

\published[Published text]{date}

\maketitle
\tableofcontents



\section{Introduction}

Studies of higher-dimension theories that involve (spontaneously) broken
symmetries and noncommutativity in the quantum case are motivated by searches
for a unified theory. Dimensional reduction of such theories is not unique and
becomes extremely involved when gravity is included. We believe that the
guiding principles for the reduction are provided by the observed (or
desirable) physical field content and by the group theoretical structure itself.

From the technical point of view, we have to extend physical fields into an
extra (internal) space with preserving the general noncommutative quantum
structure. However the development of a mechanism that permit us to display
the set of physical fields in interaction with the corresponding four
dimensional world implies that some of the original symmetries of the
higher-dimension manifold have been broken. There exist many theoretical
attempts to realize the above ideas such as string and brane theories but none
of them can be treated as the final answer: formulation of such theories
contain serious problems that are still non solved. In spite of the fact that
in these theories the solution seems to include a non-commutative
structure~\cite{Isaev:1993fs,Aschieri:2002vn}, the concrete implementation of
these symmetries in a substructure of any (super) manifold seems to be very
complicated from the technical and geometrical viewpoints.

However there exist another way to attack the unification problem that is in
the context of gauge theories of
gravity~\cite{Blagojevic:2002du,Hayashi:1981mm,Borisov:1978kw}. The first
model of gauge gravitation theory was suggested by
R.~Utiyama~\cite{Utiyama:1956sy} in 1956 just two years after the birth of
gauge theory itself. He was the first who generalized the original $SU(2)$
gauge model of Yang and Mills to an arbitrary symmetry Lie group and, in
particular, to the Lorentz group in order to describe gravity. However, he met
the problem of treating general covariant transformations and a
pseudo-Riemannian metric which had no partner in the Yang--Mills gauge theory.
To eliminate this drawback, representing a tetrad gravitational field as a
gauge field of a translation subgroup of the Poincar\'e group was attempted
because, by analogy with gauge potentials in the Yang--Mills gauge theory, the
indices of a tetrad field $\mu$ were treated as those of a translation group,
see~\cite{Blagojevic:2002du,Hayashi:1981mm,Capozziello:2011et,Hehl:1994ue,Ivanenko:1984vf,Obukhov:2006gea,Neeman:1978njh}
and references therein. Since the Poincar\'e group comes from the
Wigner--Inonu contraction of de Sitter groups $SO(2,3)$ and $SO(1,4)$ and it
is a subgroup of the conformal group, gauge theories on fibre bundles with
these structure groups were also
considered~\cite{Gotzes:1989wn,Shirafuji:1988ia,Ivanov:1981wn,Ivanov:1981wm,Leclerc:2005qc,Stelle:1979aj,Tseytlin:1981nu}%
. Because these fibre bundles fail to be natural, the lift of the group
Diff(X) of diffeomorphisms of the fiber onto the base should be
defined~\cite{Lord:1986xi,Lord:1986at}. However, these gauging approaches
contain the problem with a non-linear (translation) summand of an affine
connection being a soldering form, but neither a frame (vierbein) field nor a
tetrad field. Thus the latter doesn't have the status of a gauge
field~\cite{Greenberg:1971lectures,Kobayashi:1963book,Sardanashvily:2011rp}.
At the same time, a gauge theory in the case of spontaneous symmetry breaking
also contains classical Higgs fields, besides the gauge and matter
ones~\cite{Giachetta:2009book,Kirsch:2005st,Keyl:1991ry,Nikolova:1980ys,Sardanashvily:1992fq,Sardanashvily:2005mf,Sardanashvily:2008ke,Sardanashvily:2016kma,Trautman:1985wm}%
. Therefore, basing on the mathematical definition of a pseudo-Riemannian
metric, some authors formulated gravitation theory as a gauge theory with a
reduced Lorentz structure where a metric gravitational field is treated as a
Higgs
field~\cite{Lawson:1998yr,Sardanashvily:1980ft,Sardanashvily:1992nr,Hawking:1973uf,Sardanashvily:1984ci}%
.

The most satisfactory answer to the formulation of gravity as a gauge theory
was developed in the pure geometrical context in the works of D.V.~Volkov et
al.~\cite{Volkov:1973jd,Akulov:1975ax}; in the context of supergravity by
Arnowitt and Pran Nath~\cite{Nath:1975nj}; and finally by
Mansouri~\cite{MacDowell:1977jt} who was able to solve some of the problems
listed before by means of a principal fiber bundle imposing a condition of
orthogonality of the generators of the fiber and base manifold. Such
conditions that break the symmetry of the original group are implemented by
means of a particular choice of the metric tensor. This approach was
implemented in a supergroup structure obtaining a gauge theory of
supergravity. Note that the underlying geometry must be reductive (in the
Cartan sense) or weakly reductive in the case of supergravity.

As always, even the problem to determine which fields transform as gauge
fields and which not, as well as which fields are physical ones and which are
redundant, nonetheless remains. Also the relation between the coset
factorization (as in the case of the non-linear realization
approach~\cite{Ogievetsky:1973ik,Volkov:1973ix,Capozziello:2014gua}) and the
specific breaking of the symmetry in the pure topological theories of grand
unification (GUT) is still unclear.

\section{Coset coherent states}

Let us remind the definition of coset coherent states
\begin{equation}
\label{tag:1}H_{0}=\left\{  g\in G\mid\mathcal{U}\left(  g\right)  V_{0}
=V_{0}\right\}  \subset G.
\end{equation}
Consequently the orbit is isomorphic to the coset, e.g.
\begin{equation}
\label{tag:2}\mathcal{O}\left(  V_{0}\right)  \simeq G/H_{0}.
\end{equation}
Analogously, if we remit to the operators, e.g.
\begin{equation}
\label{tag:3}\left\vert V_{0}\right\rangle \left\langle V_{0}\right\vert
\equiv\rho_{0}%
\end{equation}
then the orbit
\begin{equation}
\label{tag:4}\mathcal{O}\left(  V_{0}\right)  \simeq G/H
\end{equation}
with
\begin{align}
\label{tag:5}H  &  = \left\{  g\in G\mid\mathcal{U}\left(  g\right)  V_{0} =
\theta V_{0}\right\} \nonumber\\
&  = \left\{  g\in G\mid\mathcal{U}\left(  g\right)  \rho_{0}\mathcal{U}%
^{\dagger}\left(  g\right)  = \rho_{0}\right\}  \subset G.
\end{align}
The orbits are identified with coset spaces of $G$ with respect to the
corresponding stability subgroups $H_{0}$ and $H$ being the vectors $V_{0}$ in
the second case defined within a phase. From the quantum viewpoint $\left\vert
V_{0}\right\rangle \in\mathcal{H}$ (the Hilbert space) and $\rho_{0}%
\in\mathcal{F}$ (the Fock space) are $V_{0}$ normalized fiducial vectors (an
embedded unit sphere in $\mathcal{H}$).

\section{Symmetry breaking mechanism: the $SO(4,2)$ case}

\subsection{General considerations}

i) Let $a,b,c=1,2,3,4,5$ and $i,j,k=1,2,3,4$ (in the six-matrix
representation) then the Lie algebra of $SO\left(  2,4\right)  $ is
\begin{align}
&  i\left[  J_{ij},J_{kl}\right]  = \eta_{ik}J_{jl}+\eta_{jl}J_{ik} -
\eta_{il}J_{jk}-\eta_{jk}J_{il},\label{tag:7}\\
&  i\left[  J_{5i},J_{jk}\right]  = \eta_{ik}J_{5j}-\eta_{ij}J_{5k}%
,\label{tag:8}\\
&  i\left[  J_{5i},J_{5j}\right]  = -J_{ij},\label{tag:9}\\
&  i\left[  J_{6a},J_{bc}\right]  = \eta_{ac}J_{6b} - \eta_{ab}J_{6c}%
,\label{tag:10}\\
&  i\left[  J_{6a},J_{6b}\right]  = -J_{ab}. \label{tag:11}%
\end{align}

ii) Identifying the first set of commutation relations~(\ref{tag:7}) as the
lie algebra of the $SO\left(  1,3\right)  $ with generators $J_{ik}=-J_{ki}$.

iii) The commutation relations~(\ref{tag:7}) plus (\ref{tag:8}) and
(\ref{tag:9}) are identified as the Lie algebra $SO\left(  2,3\right)  $ with
the additional generators $J_{5i}$ and $\eta_{ij}=\left(  1,-1,-1,-1\right)  $.

iv) The commutation relations (\ref{tag:7})--(\ref{tag:11}) is the Lie algebra
$SO\left(  2,4\right)  $ written in terms of the Lorentz group $SO\left(
1,3\right)  $ with the additional generators $J_{5i}$, $J_{6b}$, and
$J_{ab}=-J_{ba}$, where $\eta_{ab}=\left(  1,-1,-1,-1,1\right)  $. It follows
that the embedding is given by the chain $SO(1,3)\subset SO(2,3)\subset
SO(2,4)$.

\bigskip

From the six dimensional matrix representation we know from that
parameterizing the coset $\mathcal{C}=\frac{SO(2,4)}{SO(2,3)}$ and
$\mathcal{P}=\frac{SO(2,3)}{SO(1,3)}$, then any element $G$ of $SO(2,4)$ is
written as
\begin{equation}
SO(2,4)\approx\frac{SO(2,4)}{SO(2,3)}\times\frac{SO(2,3)}{SO(1,3)} \times
SO(1,3), \label{tag:12}%
\end{equation}
explicitly
\begin{align}
\label{tag:13}G  &  = e^{-iz^{a}\left(  x\right)  J_{a}}G\left(  H\right)
\nonumber\\
&  = e^{-iz^{a}\left(  x\right)  J_{a}}e^{-i\varepsilon^{k}\left(  x\right)
P_{k}}H\left(  \Lambda\right)  .
\end{align}

Consequently we have $G\left(  H\right)  :H\rightarrow G$ is an embedding of
an element of $SO(2,3)$ into $SO(2,4)$ where $J_{a}\equiv\frac{1}{\lambda
}J_{6a}$ and $H\left(  \Lambda\right)  :\Lambda\rightarrow H$ is an embedding
of an element of $SO(1,3)$ into $SO(2,3)$ where $P_{k}\equiv\frac{1}{m}J_{5k}$
as follows
\begin{equation}
G=e^{-iz^{a}\left(  x\right)  J_{a}}\underset{G\left(  H\right)  }
{\underbrace{e^{-i\varepsilon^{k}\left(  x\right)  P_{k}}\underset{H\left(
\Lambda\right)  }{\text{ }\underbrace{\left(
\begin{array}
[c]{cc}%
\fbox{$%
\begin{array}
[c]{ccccc}
&  &  &  & \\
&  &  &  & \\
&  & SO(3,1) &  & \\
&  &  &  & \\
&  &  &  &
\end{array}
$} & \mathbf{0}\\
\boldsymbol{0} & \fbox{$%
\begin{array}
[c]{ccc}
&  & \\
& I_{2x2} & \\
&  &
\end{array}
$}%
\end{array}
\right)  }}}} \label{tag:14}%
\end{equation}
then any element $G$ of $SO(2,4)$ is written as the product of an $SO(2,4)$
boost, an $ADS$ boost, and a Lorentz rotation.

\section{Goldstone fields and symmetries}

i) Our starting point is to introduce two 6-dimensional vectors $V_{1}$ and
$V_{2}$ being invariant under $SO\left(  3,1\right)  $ in a canonical form.
Explicitly
\begin{equation}
\underset{V_{1}}{\underbrace{\left(
\begin{array}
[c]{c}%
0\\
0\\
0\\
0\\
A\\
0
\end{array}
\right)  }}+\underset{V_{2}}{\underbrace{\left(
\begin{array}
[c]{c}%
0\\
0\\
0\\
0\\
0\\
-B
\end{array}
\right)  }}=\left.  \underset{V_{0}}{\underbrace{\left(
\begin{array}
[c]{c}%
0\\
0\\
0\\
0\\
A\\
-B
\end{array}
\right)  }}\right\}  invariant\text{ }under\text{ }SO\left(  3,1\right)  .
\label{tag:15}%
\end{equation}

2) Now we take an element of $Sp\left(  2\right)  \subset Mp\left(  2\right)
$ embedded in the 6-dimensional matrix representation operating over $V$ as
follows
\begin{equation}
\mathcal{M}V\rightarrow\underset{Sp\left(  2\right)  \subset Mp\left(
2\right)  }{\underbrace{\left(
\begin{array}
[c]{cccccc}%
0 & 0 & 0 & 0 & 0 & 0\\
0 & 0 & 0 & 0 & 0 & 0\\
0 & 0 & 0 & 0 & 0 & 0\\
0 & 0 & 0 & 0 & 0 & 0\\
0 & 0 & 0 & 0 & a & b\\
0 & 0 & 0 & 0 & c & d
\end{array}
\right)  }}\underset{V_{0}}{\underbrace{\left(
\begin{array}
[c]{c}%
0\\
0\\
0\\
0\\
A\\
-B
\end{array}
\right)  }}=\left(
\begin{array}
[c]{c}%
0\\
0\\
0\\
0\\
A^{\prime}\\
-B^{\prime}%
\end{array}
\right)  =V^{\prime}, \label{tag:16}%
\end{equation}
where
\begin{equation}%
\begin{array}
[c]{c}%
A^{\prime}=aA-bB,\\
-B^{\prime}=cA-dB
\end{array}
\label{tag:17}%
\end{equation}
consequently we obtain a \textit{Klauder-Perelomov generalized coherent state}
with the fiducial vector $V_{0}$.

ii) The specific task to be made by the vectors is to perform the symmetry
breakdown to $SO(3,1)$. Using the transformed vectors above ($Sp(2)\sim$
$Mp\left(  2\right)  $ CS) the symmetry of $G$ can be extended to an internal
symmetry as $SU(1,1)$ given by $\widetilde{G}$ below (note that $\left\vert
\lambda\right\vert ^{2}-\left\vert \mu\right\vert ^{2}=1)$:
\begin{align}
\widetilde{G}V^{\prime}  &  = e^{-iz^{a}\left(  x\right)  J_{a}}%
\underset{\widetilde{G}\left(  H\right)  }{\underbrace{e^{-i\varepsilon
^{k}\left(  x\right)  P_{k}}\underset{\widetilde{H}\left(  \Lambda\right)
}{\text{ }\underbrace{\left(
\begin{array}
[c]{cc}%
\fbox{$%
\begin{array}
[c]{ccccc}
&  &  &  & \\
&  &  &  & \\
&  & SO(3,1) &  & \\
&  &  &  & \\
&  &  &  &
\end{array}
$} & \mathbf{0}\\
\boldsymbol{0} & \fbox{$%
\begin{array}
[c]{cc}%
\lambda & \mu\\
\mu^{\ast} & \lambda^{\ast}%
\end{array}
$}%
\end{array}
\right)  }}}}V^{\prime}=\label{tag:18}\\
&  = e^{-iz^{a}\left(  x\right)  J_{a}}\underset{G\left(  H\right)
}{\underbrace{e^{-i\varepsilon^{k}\left(  x\right)  P_{k}}\underset{H\left(
\Lambda\right)  }{\text{ }\underbrace{\left(
\begin{array}
[c]{cc}%
\fbox{$%
\begin{array}
[c]{ccccc}
&  &  &  & \\
&  &  &  & \\
&  & SO(3,1) &  & \\
&  &  &  & \\
&  &  &  &
\end{array}
$} & \mathbf{0}\\
\boldsymbol{0} & \fbox{$%
\begin{array}
[c]{cc}%
\alpha & 0\\
0 & \beta
\end{array}
$}%
\end{array}
\right)  }}}}V_{0}=GV_{0}, \label{tag:19}%
\end{align}
\begin{equation}
\mathcal{M=}\left(
\begin{array}
[c]{cccccc}%
0 & 0 & 0 & 0 & 0 & 0\\
0 & 0 & 0 & 0 & 0 & 0\\
0 & 0 & 0 & 0 & 0 & 0\\
0 & 0 & 0 & 0 & 0 & 0\\
0 & 0 & 0 & 0 & \lambda^{\ast}\alpha & -\mu\beta\\
0 & 0 & 0 & 0 & -\mu^{\ast}\alpha & \lambda\beta
\end{array}
\right)  \label{tag:20}%
\end{equation}
and if we also ask for Det$\mathcal{M}=1$ then $\alpha\beta=1$, e.g. the
additional phase: it will bring us the 10$^{th}$ Goldstone field. The other
nine are given by $z^{a}\left(  x\right)  $ and $\varepsilon^{k}\left(
x\right)  $ ($a,b,c=1,2,3,4,5$ and $i,j,k=1,2,3,4)$ coming from the
parameterization of the cosets $\mathcal{C}=\frac{SO(2,4)}{SO(2,3)}$ and
$\mathcal{P}=\frac{SO(2,3)}{SO(1,3)}$.

\section{Invariant $SO(2,4)$ action and breakdown mechanism}

\subsection{Linear in $R^{AB}$}%

\begin{equation}
S=\int\mu_{AB}\wedge R^{AB} \label{tag:21}%
\end{equation}
in this case we note first, that the $SO(2,4)$-valuated tensor $\mu_{AB}$ acts
as multiplier in $S$ (without any role in dynamics, generally speaking).
Having this fact in mind, let us consider the following points.

i) If we have two diffeomorphic (or gauge) nonequivalent $SO(2,4)$-valuated
connections, namely $\Gamma^{AB}$ and $\widetilde{\Gamma}^{AB}$, their
difference transforms as a second rank six-tensor under the action of
$SO(2,4)$
\begin{align}
\kappa^{AB}  &  = G_{\text{ }C}^{A}G_{\text{ }D}^{B}\kappa^{CD},\label{tag:22}%
\\
\kappa^{AB}  &  \equiv\widetilde{\Gamma}^{AB}-\Gamma^{AB}. \label{tag:23}%
\end{align}

ii) If we now calculate the curvature from $\widetilde{\Gamma}^{AB}$ we
obtain
\begin{equation}
\widetilde{R}^{AB}=R^{AB}+\mathcal{D}\kappa^{AB}, \label{tag:24}%
\end{equation}
where the $SO(2,4)$ covariant derivative is defined in the usual way
\begin{equation}
\mathcal{D}\kappa^{AB}=d\kappa^{AB}+\Gamma_{\text{ }C}^{A}\wedge\kappa^{CB}
+\Gamma_{\text{ }D}^{B}\wedge\kappa^{AD}. \label{tag:25}%
\end{equation}

iii) Redefining the $SO(2,4)$ six vectors as $V_{2}^{A}\equiv\psi^{A}$ and
$V_{1}^{B}\equiv\varphi^{B}$ (in order to put all in the standard notation),
the 2-form $\kappa^{AB}$ can be constructed as
\begin{equation}
\kappa^{AB}\rightarrow\psi^{\left[  A\right.  }\varphi^{\left.  B\right]  }dU.
\label{tag:26}%
\end{equation}
Then we introduce all into the $\widetilde{R}^{AB}$ ($U$\ scalar function) and
get
\begin{align}
\widetilde{R}^{AB}  &  = R^{AB}+\mathcal{D}\left(  \psi^{\left[  A\right.  }
\varphi^{\left.  B\right]  }dU\right) \nonumber\\
&  = R^{AB}+\left(  \psi^{\left[  A\right.  }\mathcal{D}\varphi^{\left.
B\right]  }-\varphi^{\left[  A\right.  }\mathcal{D}\psi^{\left.  B\right]
}\right)  \wedge dU. \label{tag:28}%
\end{align}
The next step is to find the specific form of $\mu_{AB}$ such that
$\widetilde{\mu}_{AB}=\mu_{AB}$ (invariant under tilde transformation) in
order to make the splitting of the transformed action $\widetilde{S}$
reductive as follows.

iv) Let us define
\begin{equation}
\widetilde{\theta}^{A}=\widetilde{\mathcal{D}}\varphi^{A} \label{tag:29}%
\end{equation}
with the connection $\widetilde{\Gamma}^{AB}=\Gamma^{AB}+\kappa^{AB}$, then
\begin{align}
\label{tag:30}\widetilde{\theta}^{A}  &  = \underset{\theta^{A}}%
{\underbrace{\mathcal{D} \varphi^{A}}}+\kappa_{\text{ }B}^{A}\varphi
^{B},\nonumber\\
\widetilde{\theta}^{A}  &  = \theta^{A}+\left[  \psi^{A}\left(  \varphi
^{B}\right)  ^{2}-\varphi^{A}\left(  \psi\cdot\varphi\right)  \right]  \wedge
dU,
\end{align}
where $\left(  \varphi^{B}\right)  ^{2}=\left(  \varphi_{\text{ }B}\varphi
^{B}\right)  $ and $\left(  \psi\cdot\varphi\right)  =\psi_{B}\varphi^{B}$ etc.

In the same manner we also define
\begin{align}
\label{tag:32}\widetilde{\eta}^{A}  &  =\widetilde{\mathcal{D}}\psi
^{A},\nonumber\\
\widetilde{\eta}^{A}  &  =\eta^{A}+\left[  \psi_{2}^{A}\left(  \psi
\cdot\varphi\right)  -\varphi^{A}\left(  \psi^{B}\right)  ^{2}\right]  \wedge
dU.
\end{align}

v) To determine $\mu_{AB}$ we propose to cast it in the form%
\begin{equation}
\mu_{AB}\propto\rho_{s}\left[  a\psi^{F}\varphi^{E}\epsilon_{ABCDEF}\left(
\theta^{C}\wedge\eta^{D}+\theta^{C}\wedge\theta^{D}+\eta^{C}\wedge\eta
^{D}\right)  +b\kappa^{AB}\right]  \label{tag:34}%
\end{equation}
with $\rho_{s},a,b$ scalar functions in particular contractions of vectors and
bivectors $SO(2,4)$-valuated with $\epsilon_{ABCDEF})$ to be determined. The
behaviour under the tilde transformation is%
\begin{equation}
\label{tag:35}\widetilde{\mu}_{AB}\propto\mu_{AB}-\frac{1}{2}\rho_{s}a\psi
^{F}\varphi^{E}\epsilon_{ABEF}d\xi\wedge dU,
\end{equation}
where $\xi=\left(  \psi^{A}\right)  ^{2}\left(  \varphi^{B}\right)
^{2}-\left(  \psi\cdot\varphi\right)  ^{2}$.

vi) Finally we have to look at the behaviour of the transformed action
\begin{align}
\label{tag:36}\widetilde{S}  &  = \int\widetilde{\mu}_{AB}\wedge\widetilde
{R}^{AB}\nonumber\\
&  = S +\int\frac{1}{2}\rho_{s}a\kappa_{AB}\wedge R^{AB}\wedge d\xi+\int
\mu_{AB}\wedge\mathcal{D}\kappa^{AB}.
\end{align}
We see that till this point, the $SO(2,4)$-valuated six-vectors $\psi^{F}$ and
$\varphi^{E}$ are in principle arbitrary. However, under the conditions
discussed in the first Section the vectors go to the fiducial ones modulo a
phase. Consequently
\begin{equation}
\label{tag:37}\xi\rightarrow A^{2}B^{2}%
\end{equation}
and the bivector comes to
\begin{equation}
\label{tag:38}\kappa^{AB}\rightarrow\psi^{\left[  A\right.  }\varphi^{\left.
B\right]  }dU\rightarrow\Delta\left(  AB\right)  \epsilon^{\alpha\beta}%
=\alpha\beta AB\epsilon^{\alpha\beta}=AB\epsilon^{\alpha\beta}, \qquad
\alpha,\beta:5,6,
\end{equation}
where we define the 2nd rank antisymmetric tensor $\epsilon^{\alpha\beta}$ and%
\begin{equation}
\label{tag:39}\Delta=Det\left(
\begin{array}
[c]{cc}%
\lambda^{\ast}\alpha & -\mu\beta\\
-\mu^{\ast}\alpha & \lambda\beta
\end{array}
\right)  =\alpha\beta=1(\text{unitary transformation})
\end{equation}

Below we consider two important cases with respect to the components $m$ and
$\lambda$.

\subsection{$A=m$ and $B=\lambda$}

\begin{enumerate}
\item[1] If the coefficients $A=m$ and $B=\lambda$ play the role of
\textit{constant parameters} we have
\begin{equation}
\label{tag:40}d\xi\rightarrow d\left(  \lambda^{2}m^{2}\right)  =0
\end{equation}
and
\begin{equation}
\label{tag:41}\mathcal{D}\kappa^{AB}\rightarrow d\left(  \lambda m\right)
\epsilon^{\alpha\beta}\wedge dU=0
\end{equation}
making the original action $S$ invariant, e.g.
\begin{equation}
\label{tag:42}\left.  \widetilde{S}\right\vert _{V_{0}}=\int\widetilde{\mu
}_{AB} \wedge\widetilde{R}^{AB}=\int\mu_{AB}\wedge R^{AB}=S
\end{equation}
being $\left.  \widetilde{S}\right\vert _{V_{0}}$ the restriction of
$\widetilde{S}$ under the subspace generated by $V_{0}$ and consequently
breaking the symmetry from $SO\left(  2,4\right)  \rightarrow SO\left(
1,3\right)  $.

\item[2] The connections after the symmetry breaking (when the mentioned
conditions with $\lambda$ and $m$ constants are fulfilled) become
\begin{align}
&  \widetilde{\Gamma}^{AB}=\Gamma^{AB}+\kappa^{AB}\Rightarrow\mathrm{b.o.s.}
\rightarrow\widetilde{\Gamma}^{ij}=\Gamma^{ij};\text{ \ }\widetilde{\Gamma
}^{i5} =\Gamma^{i5},\qquad\widetilde{\Gamma}^{i6}=\Gamma^{i6},\\
&  \qquad{\mathrm{but}} \qquad\widetilde{\Gamma}^{56}=\Gamma^{56} -\left(
\lambda m\right)  dU.
\end{align}

\item[3] Vectors $\widetilde{\theta}^{A}$ and $\widetilde{\eta}^{A}$ after the
symmetry breaking and under the same conditions become
\begin{align*}
&  \widetilde{\theta}^{A}=\underset{\theta^{A}}{\underbrace{d\varphi^{A}+
\Gamma_{\text{ }C}^{A}\wedge\varphi^{C}}}+\kappa_{\text{ }B}^{A} \varphi^{B}
\Rightarrow{\mathrm{b.o.s.}},\\
&  \widetilde{\theta}^{i}=\theta^{i} = 0 + \Gamma_{\text{ }5}^{i}m +0
\Rightarrow\theta^{i}=\Gamma_{\text{ }5}^{i}m,\\
&  \widetilde{\theta}^{5}=0=0+0=0,\\
&  \widetilde{\eta}^{A}=\underset{\theta^{A}}{\underbrace{d\psi^{A}%
+\Gamma_{\text{ }C}^{A}\wedge\psi^{C}}}+\kappa_{\text{ }B}^{A}\psi^{B}
\Rightarrow{\mathrm{b.o.s.}},\\
&  \widetilde{\eta}^{i}=\eta^{i}=0-\Gamma_{\text{ }6}^{i}\lambda
+0\Rightarrow\eta^{i}=-\Gamma_{\text{ }6}^{i}\lambda,\\
&  \widetilde{\eta}^{6}=\eta^{6}=0
\end{align*}
and evidently $\mu_{i5}=\mu_{i6}=0$.

\item[4] Consequently from the last points, curvatures become
\begin{align}
R^{ij}  &  = R_{\left\{  {}\right\}  }^{ij}+m^{-2}\theta^{i}\wedge\theta^{j}+
\lambda^{-2}\eta^{i}\wedge\eta^{j},\label{tag:43}\\
R^{i5}  &  = m^{-1}\left[  \overset{D\theta^{i}}{\overbrace{d\theta^{i} +
\omega_{\ j}^{i}\wedge\theta^{j}}}+\left(  \frac{m}{\lambda}\right)  \eta
^{i}\wedge\Gamma^{65}\right]  = m^{-1}\left[  D\theta^{i}-\frac{m}{\lambda}
\eta^{i}\wedge\Gamma^{65}\right]  ,\label{tag:44}\\
R^{i6}  &  = - \lambda^{-1}\left[  D\eta^{i}-\left(  \frac{m}{\lambda}\right)
^{-1}\theta^{i}\wedge\Gamma^{56}\right]  ,\label{tag:45}\\
R^{56}  &  = d\Gamma^{56}+\left(  m\lambda\right)  ^{-1}\theta_{i}\wedge
\eta^{i},
\end{align}
where $D$ is the $SO(1,3)$ covariant derivative.

\item[5] The tensor responsible for the symmetry breaking becomes
\begin{align}
\mu_{ij}  &  = -2\rho_{s}a\lambda m\epsilon_{ijkl}\left(  \theta^{k}\wedge
\eta^{l}+\theta^{k}\wedge\theta^{l}+\eta^{k}\wedge\eta^{l}\right)
\label{tag:47}\\
\mu_{56}  &  = -\rho_{s}b\epsilon_{56}\lambda mdU. \label{tag:48}%
\end{align}

\item[6] Consequently, with all ingredients at hand, the action will be
\begin{equation}
\label{tag:49}S=\int\mu_{AB}\wedge R^{AB}=\underset{S_{1}}{\underbrace{\int
\mu_{ij}\wedge R^{ij}}}+\underset{S_{2}}{\underbrace{\int\mu_{56}\wedge
R^{56}}},
\end{equation}
where
\begin{align}
S_{1}  &  = - 2\int\rho_{s}a\epsilon_{ijkl}\left(  \theta^{k}\wedge\eta
^{l}+\theta^{k}\wedge\theta^{l} + \eta^{k}\wedge\eta^{l}\right)  \wedge\left(
\lambda mR_{\left\{  {}\right\}  }^{ij} + \frac{\lambda}{m}\theta^{i}%
\wedge\theta^{j} + \frac{m}{\lambda}\eta^{i}\wedge\eta^{j}\right) \nonumber\\
&  = - 2\int\rho_{s}a\epsilon_{ijkl}\left(  \theta^{k}\wedge\eta^{l}
\wedge\lambda mR_{\left\{  {}\right\}  }^{ij}+\theta^{k}\wedge\theta^{l}
\wedge\lambda mR_{\left\{  {}\right\}  }^{ij}+\eta^{k}\wedge\eta^{l}
\wedge\lambda mR_{\left\{  {}\right\}  }^{ij}\right) \nonumber\\
&  - 2\int\rho_{s}a\epsilon_{ijkl}\left(  \theta^{k}\wedge\eta^{l}\wedge
\frac{\lambda}{m}\theta^{i}\wedge\theta^{j}+\theta^{k}\wedge\theta^{l}
\wedge\frac{\lambda}{m}\theta^{i}\wedge\theta^{j}+\eta^{k}\wedge\eta^{l}
\wedge\frac{\lambda}{m}\theta^{i}\wedge\theta^{j}\right) \nonumber\\
&  - 2\int\rho_{s}a\epsilon_{ijkl}\left(  \theta^{k}\wedge\eta^{l}\wedge
\frac{m}{\lambda}\eta^{i}\wedge\eta^{j}+\theta^{k}\wedge\theta^{l}\wedge
\frac{m}{\lambda}\eta^{i}\wedge\eta^{j}+\eta^{k}\wedge\eta^{l}\wedge\frac
{m}{\lambda}\eta^{i}\wedge\eta^{j}\right) \nonumber
\end{align}
and
\begin{equation}
S_{2}=-\lambda m\int\rho_{s}b\epsilon_{56}\wedge\left(  d\Gamma^{56} + \left(
m\lambda\right)  ^{-1}\theta_{i}\wedge\eta^{i}\right)  .\nonumber
\end{equation}

\item[7] At this point (the mathematical justification will come later) we can
naturally associate the tetrad field with the $\theta$-form
\begin{equation}
\theta^{k}\sim e_{a}^{k}\omega^{a} \label{tag:53}%
\end{equation}
consequently a metric can be induced in $M_{4}$:
\begin{equation}
\eta_{ab}=g_{jk}e_{a}^{j}e_{b}^{k},\qquad g_{jk}=\eta_{ab}e_{j}^{a}e_{k}%
^{b},\qquad e_{a}^{k}e_{k}^{b}=\delta_{b}^{a},\qquad\mathrm{etc.},
\label{tag:54}%
\end{equation}
where $\eta_{jk}$ is the Minkowski metric. That allows us to lift up and to
lower down indices, and $\eta^{i}$ with the following symmetry typical of a
$SU\left(  2,2\right)  $ Clifford structure
\begin{align}
&  \eta^{k}\sim f_{a}^{k}\omega^{a},\label{tag:55}\\
&  e_{j}^{a}f_{a}^{k}g_{lk}=f_{lj}=-f_{jl} \label{tag:56}%
\end{align}
that consequently allows us to introduce into the model an electromagnetic
field (that will be proportional to $f_{lj})$.

\item[8] So we can re-write the action as
\begin{align}
\label{tag:57}S_{1}  &  = - 2\int\rho_{s}a\epsilon_{ijkl}\left(  \theta
^{k}\wedge\eta^{l} + \theta^{k}\wedge\theta^{l}+\eta^{k}\wedge\eta^{l}\right)
\wedge\left(  \lambda mR_{\left\{  {}\right\}  }^{ij}+\frac{\lambda}{m}%
\theta^{i} \wedge\theta^{j}+\frac{m}{\lambda}\eta^{i}\wedge\eta^{j}\right)
\nonumber\\
&  = - 2\int\rho_{s}a \biggl[ \lambda m\left(  f_{ij} R_{\left\{  {}\right\}
}^{ij} + \left(  g_{ij}+f_{i}^{k}f_{kj}\right)  R_{\left\{  {}\right\}  }%
^{ij}\right)  + \left(  \frac{\lambda}{m} + \frac{m}{\lambda}\right)
f^{kj}f_{kj}\nonumber\\
&  + \left(  \frac{\lambda}{m}\sqrt{g} + \frac{m}{\lambda}\sqrt{f}\right)
\biggr] d^{4}x.
\end{align}

\end{enumerate}

In the above expression we have taken into account the following:

i) terms $\sim\eta\wedge\eta\wedge\eta\wedge\theta$ and $\eta\wedge
\theta\wedge\theta\wedge\theta$ vanish;

ii) terms $\sim\eta\wedge\eta\wedge\theta\wedge\theta$ and $\eta\wedge
\eta\wedge\theta\wedge\theta$ lead to $\rightarrow f^{kj}f_{kj}$;

iii) term $\sim\epsilon_{ijkl}\theta^{k}\wedge\eta^{l}\wedge R_{\left\{
{}\right\}  }^{ij}$ leads $\rightarrow f_{ij}R_{\left\{  {}\right\}  }^{ij}$
picking the antisymmetric part of the generalized Ricci tensor (containing torsion);

iv) term $\sim\epsilon_{ijkl}\left(  \theta^{k}\wedge\theta^{l}+\eta^{k}
\wedge\eta^{l}\right)  R_{\left\{  {}\right\}  }^{ij}$ leads to $\rightarrow
\left(  g_{ij}+f_{i}^{k}f_{kj}\right)  R_{\left\{  {}\right\}  }^{ij}$ picking
the symmetric part of the generalized Ricci tensor (containing
Einstein-Hilbert plus quadratic torsion term);

v) terms $\sim\eta\wedge\eta\wedge\eta\wedge\eta$ and $\theta\wedge
\theta\wedge\theta\wedge\theta$ lead to the volume elements $\sqrt{f}$ and
$\sqrt{g}$, respectively, where we defined as usual $g\equiv Det\left(
g_{lk}\right)  $ and $f\equiv Det\left(  f_{lk}\right)  =\left(  f_{lk}^{\ast
}f^{lk}\right)  ^{2}$.

\subsection{$A=m\left(  x\right)  $ and $B=\lambda\left(  x\right)  :$
spontaneous subspace}

If the coefficients $A=m\left(  x\right)  $ and $B=\lambda\left(  x\right)  $
are not \textit{constant} but functions of coordinates we have
\begin{equation}
\label{tag:58}d\xi\rightarrow d\left(  \lambda^{2}m^{2}\right)  =2d\left(
\lambda m\right)
\end{equation}
and
\begin{equation}
\label{tag:59}\mathcal{D}\kappa^{AB}\rightarrow d\left(  \lambda m\right)
\epsilon^{\alpha\beta}\wedge dU.
\end{equation}
Consequently from the following explicit computations
\begin{align}
\label{tag:60}\widetilde{S}  &  = \int\widetilde{\mu}_{AB}\wedge\widetilde
{R}^{AB}\\
&  = S+\int\frac{1}{2}\rho_{s}a\kappa_{AB}\wedge R^{AB}\wedge d\xi+ \int
\mu_{AB}\wedge\mathcal{D}\kappa^{AB}\nonumber\\
&  = S-\int\frac{1}{2}\rho_{s}aR^{AB}\wedge\kappa_{AB}\wedge d\xi+\int\mu
_{AB}\wedge\mathcal{D}\kappa^{AB}\nonumber\\
&  = S-\int\frac{1}{2}\rho_{s}aR_{\alpha\beta}\epsilon^{\alpha\beta}\lambda
mdU\wedge2d\left(  \lambda m\right)  +\int\mu_{\alpha\beta}\epsilon
^{\alpha\beta}d\left(  \lambda m\right)  \wedge dU\nonumber\\
&  = S+\int\frac{1}{2}\rho_{s}aR_{\alpha\beta}\epsilon^{\alpha\beta}\lambda
m2d\left(  \lambda m\right)  \wedge dU+\int\mu_{\alpha\beta}\epsilon
^{\alpha\beta}d\left(  \lambda m\right)  \wedge dU,\nonumber\\
\widetilde{S}  &  = S+\int\left[  \mu_{\alpha\beta}+\rho_{s}aR_{\alpha\beta
}\lambda m\right]  \epsilon^{\alpha\beta}d\left(  \lambda m\right)  \wedge
dU.\nonumber
\end{align}
we obtain the required condition:
\begin{align}
\widetilde{S}  &  = S \qquad\mathrm{if}\nonumber\\
\mu_{\alpha\beta}  &  = -\rho_{s}aR_{\alpha\beta}\lambda m, \label{tag:62}%
\end{align}
then we see that $\mu_{AB}$ takes the place of an induced metric and it is
proportional to the curvature
\begin{align}
&  R_{\alpha\beta} = \Lambda\mu_{\alpha\beta}\label{tag:63}\\
&  \mathrm{with} \qquad\Lambda= -\left(  \rho_{s}a\lambda m\right)  ^{-1}.
\label{tag:64}%
\end{align}
Note that we have now a four-dimensional space-time plus the above "internal"
space of a constant curvature. This point is very important as a new
compactification-like mechanism.

\section{Supergravity as a gauge theory and topological QFT}

In previous works~\cite{diegofop}\cite{diegoplb} we have shown, by means of a
toy model, that there exists a supersymmetric analog of the above symmetry
breaking mechanism coming from the topological QFT. Here we recall some of the
above ideas in order to see clearly the analogy between the group structures
of the simplest supersymmetric case, $Osp\left(  4\right)  $, and of the
classical conformal group $SO\left(  2,4\right)  $.

The starting point is the super $SL(2C)$ superalgebra (strictly speaking
$Osp(4)$)
\begin{align}
&  \lbrack M_{AB},M_{CD}] = \epsilon_{C} \left(  _{A}M_{B}\right)  _{D} +
\epsilon_{D}\left(  _{A}M_{B}\right)  _{C},\nonumber\\
&  \lbrack M_{AB},Q_{C}] = \epsilon_{C}\left(  _{A}Q_{B}\right)  ,
\qquad\left\{  Q_{A},Q_{B}\right\}  = 2M_{AB}. \label{tag:18a}%
\end{align}
Here the indices $A,B,C...$ stay for $\alpha,\beta,\gamma...\left(
\overset{.}{\alpha},\overset{.}{\beta},\overset{.}{\gamma}...\right)  $ spinor
indices: $\alpha,\beta\left(  \overset{.}{\alpha},\overset{.}{\beta}\right)
=1,2\left(  \overset{.}{1},\overset{.}{2}\right)  $ in the Van~der~Werden
spinor notation. We define the superconnection $A$ due the following
"gauging"
\begin{equation}
A^{p}T_{p} \equiv\omega^{\alpha\overset{.}{\beta}}M_{\alpha\overset{.}{\beta}}
+ \omega^{\alpha\beta}M_{\alpha\beta}+\omega^{\overset{.}{\alpha}\overset
{.}{\beta}} M_{\overset{.}{\alpha}\overset{.}{\beta}}+\omega^{\alpha}%
Q_{\alpha} - \omega^{\overset{.}{\alpha}}\overline{Q}_{\overset{.}{\alpha}},
\label{tag:19a}%
\end{equation}
where $\left(  \omega M\right)  $ defines a ten-dimensional bosonic
manifold\footnotetext[3]{Corresponding to the number of generators of
$SO\left(  4,1\right)  $ or $SO\left(  3,2\right)  $ that define the group
manifold} and $p\equiv$multi-index, as usual. Analogically the super-curvature
is defined by $F\equiv F^{p}T_{p}$ with the following detailed structure
\begin{align}
F\left(  M\right)  ^{AB}  &  = d\omega^{AB}+\omega_{\ C}^{A}\wedge\omega^{CB}
+\omega^{A}\wedge\omega^{B},\label{tag:20a}\\
F\left(  Q\right)  ^{A}  &  = d\omega^{A}+\omega_{\ C}^{A}\wedge\omega^{C}.
\label{tag:21a}%
\end{align}
From (\ref{tag:19a}) it is easy to see that there are a bosonic part and a
fermionic one associated with the even and odd generators of the superalgebra.
Our proposal for the "toy" action was (as before for $SO(2,4)$) as follows:
\begin{equation}
S=\int F^{p}\wedge\mu_{p}, \label{tag:22a}%
\end{equation}
where the tensor $\mu_{p}$ (that plays the role of a $Osp\left(  4\right)  $
diagonal metric as in the Mansouri proposal) is \textit{defined} as
\begin{equation}%
\begin{array}
[c]{ccc}%
\mu_{\alpha\overset{.}{\beta}}=\zeta_{\alpha}\wedge\overline{\zeta}%
_{\overset{.}{\beta}} & \mu_{\alpha\beta}=\zeta_{\alpha}\wedge\zeta_{\beta} &
\mu_{\alpha}=\nu\zeta_{\alpha}%
\end{array}
etc. \label{tag:23a}%
\end{equation}
with $\zeta_{\alpha}\left(  \overline{\zeta}_{\overset{.}{\beta}}\right)  $
anti-commuting spinors (suitable basis\footnotetext[4]{In general this tensor
has the same structure as the Cartan-Killing metric of the group under
consideration.}) and $\nu$ the parameter of the breaking of super $SL(2C)$
($Osp\left(  4\right)  )$ to $SL(2C)$ symmetry of $\mu_{p}$. Note that the
introduction of the parameter $\nu$ means that we do not take care of the
particular dynamics to break the symmetry.

In order to obtain dynamical equations of the theory, we proceed to perform
variation of the proposed action (\ref{tag:22a})
\begin{align}
\label{tag:24a}\delta S  &  = \int\delta F^{p}\wedge\mu_{p}+F^{p}\wedge
\delta\mu_{p}\nonumber\\
&  = \int d_{A}\mu_{p}\wedge\delta A^{p}+F^{p}\wedge\delta\mu_{p},
\end{align}
where $d_{A}$ is the exterior derivative with respect to the super-$SL\left(
2C\right)  $ connection and $\delta F=d_{A}\delta A$ have been used. Then, as
the result, the dynamics is described by
\begin{equation}
d_{A}\mu=0, \qquad F=0. \label{tag:25a}%
\end{equation}
The fist equation claims that $\mu$ is covariantly constant with respect to
the super $SL\left(  2C\right)  $ connection. This fact will be very important
when the super $SL\left(  2C\right)  $ symmetry breaks down to $SL\left(
2C\right)  $ because $d_{A}\mu=d_{A}\mu_{AB}+d_{A}\mu_{A}=0$, a soldering form
will appear. The second equation gives the condition for a super Cartan
connection $A=\omega^{AB}+\omega^{A}$ to be flat, as it is easy to see from
the reductive components of above expressions
\begin{align}
F\left(  M\right)  ^{AB}  &  = R^{AB}+\omega^{A}\wedge\omega^{B}%
=0,\label{tag:26a}\\
F\left(  Q\right)  ^{A}  &  = d\omega^{A} + \omega_{\ C}^{A}\wedge\omega^{C} =
d_{\omega}\omega^{A}=0,\nonumber
\end{align}
where now $d_{\omega}$ is the exterior derivative with respect to the
$SL\left(  2C\right)  $ connection and $R^{AB}\equiv d\omega^{AB}+\omega
_{\ C}^{A}\wedge\omega^{CB}$ is the $SL\left(  2C\right)  $ curvature. Then
\begin{equation}
F=0\Leftrightarrow R^{AB}+\omega^{A}\wedge\omega^{B}=0\qquad\text{ and }\qquad
d_{\omega}\omega^{A}=0 \label{tag:27a}%
\end{equation}
the second condition says that the $SL\left(  2C\right)  $ connection is
super-torsion free. The first doesn't say that the $SL\left(  2C\right)  $
connection is flat, but it claims that it is homogeneous with a cosmological
constant related to the explicit structure of the Cartan forms $\omega^{A}$,
as we will see when the super $SL\left(  2C\right)  $ action is reduced to the
Volkov-Pashnev model~\cite{Volkov:1980mg}.

\subsection{The geometrical reduction: extended symplectic super-metrics}

\subsubsection{Example: Volkov-Pashnev metric}

The super-metric under consideration, proposed by Volkov and Pashnev
in~\cite{Volkov:1980mg}, is the simplest example of symplectic (super) metrics
induced by the symmetry breaking from a pure topological first order action.
It can be obtained from the $Osp\left(  4\right)  \left(  superSL\left(
2C\right)  \right)  $ action via the following procedure.

i) The In\"{o}nu-Wigner contraction~\cite{Inonu:1953sp} in order to pass from
$SL\left(  2C\right)  $ to the super-Poincare algebra (corresponding to the
original symmetry of the model of
refs.~\cite{CiriloLombardo:2007zk,Volkov:1980mg}) then, the even part of the
curvature is split into a $\mathbb{R}^{3,1}$ part $R^{\alpha\overset{.}{\beta
}}$ and a $SO\left(  3,1\right)  $ part $R^{\alpha\beta}\left(  R^{\overset
{.}{\alpha}\overset{.}{\beta}}\right)  $ associated with the remaining six
generators of the original five dimensional $SL\left(  2C\right)  $ group.
This fact is easily realized by knowing that the underlying geometry is
reductive: $SL\left(  2C\right)  \sim SO\left(  4,1\right)  $ $\rightarrow$
$SO\left(  3,1\right)  $ $+$ $\mathbb{R}^{3,1}$. Than we rewrite the
superalgebra (\ref{tag:18a}) as
\begin{equation}%
\begin{array}
[c]{ccc}%
\lbrack M,M]\sim M & [M,\Pi]\sim\Pi & [\Pi,\Pi]\sim M\\
\lbrack M,S]\sim S & [\Pi,S]\sim S & \left\{  S,S\right\}  \sim M+\Pi
\end{array}
\label{tag:28a}%
\end{equation}
with $\Pi\sim M_{\alpha\overset{.}{\beta}}$, $M\sim M_{\alpha\beta}\left(
M_{\overset{.}{\alpha}\overset{.}{\beta}}\right)  $, and re-scale $m^{2}\Pi=P$
and $mS=Q$. In the limit $m\rightarrow0$, one recovers the super Poincare
algebra. Note that one does not re-scale $M$ since one wants to keep
$[M,M]\sim M$ Lorentz algebra, that also is a symmetry of (\ref{tag:1}).

ii) The \textit{spontaneous} breaking of the super $SL\left(  2C\right)  $
down to the $SL\left(  2C\right)  $ symmetry of $\mu_{p}$ (e.g. $\nu
\rightarrow0$ in $\mu_{p}$) of such a manner that the even part of the super
$SL\left(  2C\right)  $ action $F\left(  M\right)  ^{AB}$ remains.

After these evaluations, it has been explicitly realized that the even part of
the original super $SL\left(  2C\right)  $ action (now a super-Poincare
invariant) can be related with the original metric~(\ref{tag:1}) as follows:
\begin{equation}
R\left(  M\right)  +R\left(  P\right)  +\omega^{\alpha}\omega_{\alpha}
-\omega^{\overset{.}{\alpha}}\omega_{\overset{.}{\alpha}}\rightarrow
\omega^{\mu}\omega_{\mu}+\mathbf{a}\omega^{\alpha}\omega_{\alpha}%
-\mathbf{a}^{\ast}\omega^{\overset{.}{\alpha}}\omega_{\overset{.}{\alpha}}%
\mid_{VP}. \label{tag:29a}%
\end{equation}
Note that there is mapping $R\left(  M\right)  + R\left(  P\right)
\rightarrow\omega^{\mu}\omega_{\mu}\mid_{VP}$ that is well defined and can be
realized in different forms, and the map of interest here $\omega^{\alpha
}\omega_{\alpha}-\omega^{\overset{.}{\alpha}}\omega_{\overset{.}{\alpha}}$
$\rightarrow$ $\mathbf{a}\omega^{\alpha}\omega_{\alpha}-\mathbf{a}^{\ast
}\omega^{\overset{.}{\alpha}} \omega_{\overset{.}{\alpha}}\mid_{VP}$ that
associate the Cartan forms of the original super $SL\left(  2C\right)  $
action (\ref{tag:22a}) with the Cartan forms of the Volkov-Pashnev supermodel:
$\omega^{\alpha}=\left(  \mathbf{a}\right)  ^{1/2}\omega^{\alpha}\mid_{VP}$,
$\omega^{\overset{.}{\alpha}}=\left(  \mathbf{a}^{\ast}\right)  ^{1/2}
\omega^{\overset{.}{\alpha}}\mid_{VP}$. Then, the origin of the coefficients
$\mathbf{a}$ and $\mathbf{a}^{\ast}$ becomes clear from the geometrical point
of view.

From the first condition in~(\ref{tag:27a}) and the association~(\ref{tag:29a}%
) it is not difficult to see that, as in the case of the space-time
cosmological constant $\Lambda:R=\frac{\Lambda}{3}e\wedge e\ \left(  e\equiv
space-time\ tetrad\right)  $, there is a cosmological term from the superspace
related to the complex parameters $\mathbf{a}$ and $\mathbf{a}^{\ast}$:
$R=-\left(  \mathbf{a}\omega^{\alpha}\omega_{\alpha}-\mathbf{a}^{\ast}
\omega^{\overset{.}{\alpha}}\omega_{\overset{.}{\alpha}}\right)  $ and it is
easy to see from the minus sign in above expression, why for
supersymmetric\ (supergravity) models it is more natural to use $SO\left(
3,2\right)  $ instead of $SO\left(  4,1\right)  $.

Note that the role of the associated spinorial action in~(\ref{tag:22a}) is
constrained by the nature of $\nu\zeta_{\alpha}$ in $\mu_{p}$ as follows.

i) If they are of the same nature of the $\omega^{\alpha}$, this term is a
total derivative and has not influence onto the equations of motion, then the
action proposed by Volkov and Pashnev in~\cite{Volkov:1980mg} has the correct
fermionic form.

ii) If they are not of the same $SL\left(  2C\right)  $ invariance that the
$\omega^{\alpha}$, the symmetry of the original model is modified. In this
direction a relativistic supersymmetric model for particles was proposed in
ref.~\cite{deAzcarraga:1984hf} considering an N-extended Minkowsky superspace
and introducing central charges to the superalgebra. Hence the underlying
rigid symmetry gets enlarged to N-extended super-Poincare algebra. Considering
for our case similar superextension that in ref.~\cite{deAzcarraga:1984hf} we
can introduce the following new action
\begin{align}
S  &  =-m\int_{\tau1}^{\tau2}d\tau\sqrt{\overset{\circ}{\omega_{\mu}}%
\overset{\circ}{\omega^{\mu}}+a\overset{.}{\theta}^{\alpha}\overset{.}{\theta
}_{\alpha}-a^{\ast}\overset{.}{\overline{\theta}}^{\overset{.}{\alpha}%
}\overset{.}{\overline{\theta}}_{\overset{.}{\alpha}}+i(\theta^{\alpha
i}A_{ij}\overset{.}{\theta}_{\alpha}^{j}-\overline{\theta}^{\overset{.}%
{\alpha}i}A_{ij}\overset{.}{\overline{\theta}}_{\overset{.}{\alpha}}^{j}%
)}\nonumber\label{tag:30a}\\
&  =\int_{\tau1}^{\tau2}d\tau L\left(  x,\theta,\overline{\theta}\right)
\end{align}
that is the super-extended version of the superparticle model proposed
in~\cite{Volkov:1980mg} with the addition of a first-order fermionic part. The
matrix tensor $A_{ij}$ introduce the symplectic structure of such manner that
now $\zeta_{\alpha i}\sim A_{ij}\theta_{\alpha}^{j}$ is not covariantly
constant under $d_{\omega}$. Note that the "Dirac-like" fermionic part is
obviously \textit{under} the square root because it is a part of the full
curvature, fact that was not advertised by the authors
in~\cite{deAzcarraga:1984hf} (see also~\cite{Sardanashvily:2005mf}) that
doesn't take into account the geometrical origin of the action. An interesting
point is to perform the same quantization as in the first part of the research
given in \cite{CiriloLombardo:2007zk} in order to obtain and compare the
spectrum of physical states with the one obtained in
ref.~\cite{deAzcarraga:1984hf}. This issue will be presented
elsewhere~\cite{Cirilo-Lombardo:2017Arbuzov}.

The spontaneous symmetry breaking happens here because the parameter doesn't
have any dynamics. But this doesn't happen in the nonlinear realization
approach where the parameters have a particular dynamics associated with the
space-time coordinates.

\section{Quadratic in $R^{AB}$}

The previous action, linear in the generalized curvature, has some drawbacks
that make necessary introduction of additional "subsidiary conditions" due to
the fact that the curvatures $R^{i5}$ and $R^{i6}$ don't play any role in the
linear/first order action. Such curvatures have a very important information
about the dynamics of $\theta$ and $\eta$ fields. In order to simplify the
equations of motion we define%
\begin{align}
\Gamma^{56}  &  \equiv A,\label{tag:65}\\
m^{-1}\theta^{i}  &  \equiv\widetilde{\theta}^{i},\label{tag:66}\\
\lambda^{-1}\eta^{i}  &  \equiv\widetilde{\eta}^{i}, \label{tag:67}%
\end{align}
and as always
\begin{equation}
R^{ij}=R_{\left\{  {}\right\}  }^{ij}+m^{-2}\theta^{i}\wedge\theta^{j} +
\lambda^{-2}\eta^{i}\wedge\eta^{j} \label{tag:68}%
\end{equation}
with the $SO\left(  1,3\right)  $ curvature $R_{\left\{  {}\right\}  }%
^{ij}=d\omega^{ij}+\omega_{\text{ }\lambda}^{i}\wedge\omega^{\lambda j}$.
Consequently from the quadratic Lagrangian density
\begin{equation}
\label{tag:69}S=\int R_{AB}\wedge R^{AB}%
\end{equation}
we obtain the following equations of motion:
\begin{align}
\label{tag:70} &  \frac{\delta\left(  R_{AB}\wedge R^{AB}\right)  }%
{\delta\theta^{i}} \rightarrow D \left(  D\widetilde{\theta}_{j}\right)  +
2R_{ij}\wedge\widetilde{\theta}^{i} - \widetilde{\theta}^{i}\wedge
\widetilde{\eta}_{i}\wedge\widetilde{\eta}_{j} + \widetilde{\theta}_{j}\wedge
A\wedge A=0,\\
&  \frac{\delta\left(  R_{AB}\wedge R^{AB}\right)  }{\delta\eta^{i}}
\rightarrow D\left(  D\widetilde{\eta}_{j}\right)  + 2R_{jk}\wedge
\widetilde{\eta}^{k} - \widetilde{\theta}^{i}\wedge\widetilde{\eta}_{i}%
\wedge\widetilde{\theta}_{j}+\widetilde{\eta}_{j}\wedge A\wedge
A=0,\label{tag:71}\\
&  \frac{\delta\left(  R_{AB}\wedge R^{AB}\right)  }{\delta\Gamma^{56}}
\rightarrow\widetilde{\theta}^{i}\wedge\widetilde{\theta}_{i} = \widetilde
{\eta}^{i}\wedge\widetilde{\eta}_{i},\label{tag:72}\\
&  \frac{\delta\left(  R_{AB}\wedge R^{AB}\right)  }{\delta\omega_{j}^{i}}
\rightarrow- DR_{kl}+D\widetilde{\theta}_{k}\wedge\widetilde{\theta}%
_{l}+D\widetilde{\eta}_{k}\wedge\widetilde{\eta}_{l}+\widetilde{\theta}%
_{k}\wedge\widetilde{\eta}_{l}\wedge A=0. \label{tag:73}%
\end{align}

\subsection{Maxwell equations and the electromagnetic field}

As we claimed before we can identify
\begin{align}
\theta^{i}  &  \equiv e_{\mu}^{i}dx^{\mu},\label{tag:74}\\
\eta^{i}  &  \equiv f_{\mu}^{i}dx^{\mu} \label{tag:75}%
\end{align}
with the symmetries
\begin{equation}
\label{tag:76}e_{\mu}^{i}e_{i}^{\nu}=\delta_{\mu}^{\nu},e_{\mu}^{i}e_{i\nu}
=g_{\mu\nu}=g_{\nu\mu}%
\end{equation}
and%
\begin{equation}
f_{\mu}^{i}f_{i}^{\nu}=\delta_{\mu}^{\nu},\qquad e_{i\nu}f_{\mu}^{i}=f_{\mu
\nu}=-f_{\nu\mu} \label{tag:77}%
\end{equation}
such that the geometrical (Bianchi) condition
\begin{equation}
\nabla_{\left[  \rho\right.  }f_{\left.  \mu\nu\right]  } =\nabla_{\rho}%
^{\ast}f^{\rho\nu}=0 \label{tag:78}%
\end{equation}
or in the language of differential forms
\begin{equation}
\label{tag:79}D\left(  \widetilde{\theta}^{i}\wedge\widetilde{\eta}%
_{i}\right)  =0
\end{equation}
holds, thus the curvatures $R^{i6}$ and $R^{i5}$ are enforced to be null. And
conversely if $R^{i6}$ and $R^{i5}$ are zero then $D\left(  \widetilde{\theta
}^{i}\wedge\widetilde{\eta}_{i}\right)  =0$ or equivalently $\nabla_{\left[
\rho\right.  }f_{\left.  \mu\nu\right]  }=\nabla_{\rho}^{\ast}f^{\rho\nu}=0$.

\begin{proof}
From expressions (\ref{tag:44},\ref{tag:45}), namely: $R^{i5}=\left[
D\widetilde{\theta}^{i}-\widetilde{\eta}^{i}\wedge\Gamma^{65}\right]  $ and
$R^{i6}=\left[  -D\widetilde{\eta}^{i}+\widetilde{\theta}^{i}\wedge\Gamma
^{56}\right]  $ we make
\begin{align}
&  R^{i5}\wedge\widetilde{\eta}_{i}+\widetilde{\theta}_{i}\wedge R^{i6} =
D\left(  \widetilde{\theta}^{i}\wedge\widetilde{\eta}_{i}\right)  +\left(
\widetilde{\eta}^{i}\wedge\Gamma^{56}\right)  \wedge\widetilde{\eta}_{i} +
\widetilde{\theta}_{i}\wedge\left(  \widetilde{\theta}^{i}\wedge\Gamma
^{56}\right)  ,\label{tag:80}\\
&  R^{i5}\wedge\widetilde{\eta}_{i}+\widetilde{\theta}_{i}\wedge R^{i6} =
D\left(  \widetilde{\theta}^{i}\wedge\widetilde{\eta}_{i}\right)  .
\label{tag:81}%
\end{align}
In the last line we used the constraint given by eq.~(\ref{tag:72})
Consequently if $R^{i6}$ and $R^{i5}$ are zero, then $D\left(  \widetilde
{\theta}^{i}\wedge\widetilde{\eta}_{i}\right)  =0$ or equivalently
$\nabla_{\left[  \rho\right.  }f_{\left.  \mu\nu\right]  }=\nabla_{\rho}%
^{\ast}f^{\rho\nu}=0$ and vice versa.

\begin{corollary}
Note that the vanishing of the $R^{56}$ curvature (that transforms as a
Lorentz scalar) does not modify the equation of motion for $\Gamma^{56}$ and
simultaneously defines the electromagnetic field as
\begin{align}
&  R^{56} = d\Gamma^{56}+\left(  m\lambda\right)  ^{-1}\theta_{i}\wedge
\eta^{i}=0,\label{tag:82}\\
&  \Rightarrow dA - F =0. \label{tag:83}%
\end{align}

\end{corollary}
\end{proof}

\subsection{Equations of motion in components and symmetries}

Let us define
\begin{align}
&  R_{\left\{  {}\right\}  \mu\nu}^{ij}=\partial_{\mu}\omega_{\nu}^{ij} -
\partial_{\nu}\omega_{\mu}^{ij}+\omega_{\mu k}^{i}\omega_{\nu}^{kj} -
\omega_{\mu}^{kj}\omega_{\nu k}^{i},\label{tag:84}\\
&  T_{\mu\nu}^{i}=\partial_{\mu}e_{\nu}^{i}-\partial_{\nu}e_{\mu}^{i} +
\omega_{\mu\ k}^{i}e_{\nu}^{k}-\omega_{\nu\text{ }k}^{\text{ }i}e_{\mu}%
^{k},\label{tag:85}\\
&  S_{\mu\nu}^{i} = \partial_{\mu}f_{\nu}^{i}-\partial_{\nu}f_{\mu}^{i} +
\omega_{\mu\ k}^{i}f_{\nu}^{k}-\omega_{\nu\text{ }k}^{\text{ }i}f_{\mu}^{k}.
\label{tag:86}%
\end{align}
Note that $S_{\mu\nu}^{i}$ is a totally antisymmetric torsion field due the
symmetry of $f_{\nu}^{i}dx^{\nu}\equiv\eta^{i}$. Consequently the equations of
motion in components become
\begin{align}
&  \nabla_{\mu}\left[  \sqrt{\left\vert g\right\vert }R^{ij\mu\nu}\right]
+\sqrt{\left\vert g\right\vert }\left(  -m^{-2}T^{ji\nu}+\lambda^{-2}S^{ji\nu
}\right)  -\sqrt{\left\vert g\right\vert } \left(  \lambda m\right)
^{-1}f^{\left[  i\right.  \nu}A^{\left.  i\right]  }=0,\nonumber\\
&  \nabla_{\mu}\left[  \sqrt{\left\vert g\right\vert }\left(  R_{\left\{
{}\right\}  }^{ij\mu\nu}-m^{-2}e^{\left[  i\right.  \mu}e^{\left.  j\right]
\nu} + \lambda^{-2}f^{\left[  i\right.  \mu}f^{\left.  j\right]  \nu}\right)
\right] \nonumber\\
&  \quad+ \sqrt{\left\vert g\right\vert }\left(  -m^{-2}T^{ji\nu} +
\lambda^{-2}S^{ji\nu}\right)  -\sqrt{\left\vert g\right\vert } \left(  \lambda
m\right)  ^{-1}f^{\left[  i\right.  \nu}A^{\left.  i\right]  }=0,\nonumber\\
&  \nabla_{\mu}\left(  \sqrt{\left\vert g\right\vert }T^{j\mu v}\right)  +
\sqrt{\left\vert g\right\vert }\left(  R_{\left\{  {}\right\}  }^{j\nu} -
m^{-2}e^{j\nu}+A^{i}A^{\nu}\right)  =0,\nonumber\\
&  \nabla_{\mu}\left(  \sqrt{\left\vert g\right\vert }S^{j\mu i}\right)
+\sqrt{\left\vert g\right\vert }\left(  R_{\left\{  {}\right\}  }^{ij}
-\lambda^{-2}f^{ij}+A^{\left[  i\right.  }A^{\left.  j\right]  }\right)
=0,\nonumber\\
&  \nabla_{\left[  \mu\right.  }A_{\left.  \nu\right]  }=F_{\mu\nu}=\left(
\lambda m\right)  ^{-1}F_{\mu\nu},\nonumber\\
&  \nabla_{\left[  \rho\right.  }F_{\left.  \mu\nu\right]  }=0.
\end{align}

\section{Nonlinear realizations viewpoint}

Note that in our case eqs.~(\ref{tag:74},\ref{tag:75}) identify $\theta
^{i}\sim e^{i}$ and $\eta^{i}\sim f^{i}$ making the table below completely
clear. Note that $\Gamma^{65}$ is identified with the $\mathbf{g}$ of
E.~Ivanov and J.~Niederle~\cite{Ivanov:1981wn,Ivanov:1981wm}.%

\[%
\begin{tabular}
[c]{|l|l|l|}\hline
& this work & \cite{Ivanov:1981wn,Ivanov:1981wm}\\\hline
$R^{ij}$ & $R_{\left\{  {}\right\}  }^{ij}+m^{-2}\theta^{i}\wedge\theta^{j}
+\lambda^{-2}\eta^{i}\wedge\eta^{j}$ & $R_{\left\{  {}\right\}  }^{ij} +
4ge^{i}\wedge f^{j}$\\\hline
$R^{i5}$ & $m^{-1}\left[  D\theta^{i}-\frac{m}{\lambda}\eta^{i}\wedge
\Gamma^{65}\right]  $ & $De^{i}+2ge^{i}\wedge\mathbf{g}$\\\hline
$R^{i6}$ & $-\lambda^{-1}\left[  D\eta^{i}-\left(  \frac{m}{\lambda}\right)
^{-1}\theta^{i}\wedge\Gamma^{56}\right]  $ & $Df^{i}-2gf^{i}\wedge\mathbf{g}%
$\\\hline
$R^{56}$ & $d\Gamma^{56}+\left(  m\lambda\right)  ^{-1}\theta_{i}\wedge
\eta^{i}$ & $d\mathbf{g}+4ge_{i}\wedge f^{i}$\\\hline
DS/ADS\ reduction & Yes & No\\\hline
\end{tabular}
\ \ \ \
\]

Algebra and transformations in the case of the work of Ivanov and Niederle are
different due different definitions of the generators of the $SO(2,4)$
algebra, however the meaning of \textbf{g} which is associated to the
connection $\Gamma^{65}$ remains obscure for us because of the second Cartan
structure equations $R^{i5}$ and $R^{i6}$. Note that, although the group
theoretical viewpoint in the case of the simultaneous nonlinear realization of
the affine and conformal group \cite{Borisov:1974bn} to obtain Einstein
gravity are more or less clear, the pure geometrical picture is still hard to
recognize due the factorization problem and the orthogonality between coset
elements and the corresponding elements of the stability subgroup.

\section{Discussion}

In this work we introduced two geometrical models: one linear and another one
quadratic in curvature. Both models are based on the $SO(2,4)$ group.
Dynamical breaking of this symmetry was considered. In both cases we
introduced coherent states of the Klauder-Perelomov type, which as defined by
the action of a group (generally a Lie group) are invariant with respect to
the stability subgroup of the corresponding coset being related to the
possible extension of the connection which maintains the proposed action invariant.

The linear action, unlike the cases of West, Kerrick or even McDowell and
Mansouri~\cite{MacDowell:1977jt}, uses a symmetry breaking tensor that is
dynamic and unrelated to a particular metric. Such a tensor depends on the
introduced vectors (i.e. the coherent states) that intervene in the extension
of the permissible symmetries of the original connection. Only some components
of the curvature, defined by the second structure equation of Cartan, are
involved in the action, leaving the remaining ones as a system of independent
or ignorable equations in the final dynamics. The quadratic action, however,
is independent of any additional structure or geometric artifacts and all the
curvatures (e.g. all the geometrical equations for the fields) play a role in
the final action (Lagrangian of the theory).

With regard to the parameters that come into play $\mathbf{\lambda}$ and $m$
(they play the role of a cosmological constant and a mass, respectively) we
saw that in the case of linear action if they are taken dependent on the
coordinates and under the conditions of the action invariance, a new
spontaneous compactification mechanism is defined in the subspace invariant
under the stability subgroup.

Following this line of research with respect to possible physical
applications, we are going to consider scenarios of the Grand Unified Theory,
derivation of the symmetries of the Standard Model together with the
gravitational ones. The general aim is to obtain in a precisely established
way the underlying fundamental theory. This will be important, in particular,
to solve the problem of hierarchies and fundamental constants, the masses of
physical states, and their interaction.

\section{Acknowledgements}

D.J. Cirilo-Lombardo is grateful to the Bogoliubov Laboratory of Theoretical
Physics-JINR and CONICET-ARGENTINA for financial support.

\end{document}